\documentclass[conference]{IEEEtran}
\usepackage{amsmath,amssymb,epsfig,xcolor} 
\usepackage{tabularx,multirow,array}
\usepackage{times}
\usepackage{booktabs}
\usepackage{multirow}
\usepackage{soul}
\usepackage{enumerate}
\usepackage[countmax]{subfloat}
\usepackage{algpseudocode}
\usepackage{algorithm}
\usepackage{subfigure}
\usepackage{threeparttable}
\usepackage{cite}
\usepackage{url}

\begin{document}

\title{Balancing Data Security and Blocking Performance with Spectrum Randomization in Optical Networks}

\author{
    \IEEEauthorblockN{Sandeep Kumar~Singh, Wolfgang Bziuk, and Admela~Jukan}
    \IEEEauthorblockA{Technische Universit\"at Carolo-Wilhelmina zu Braunschweig, Germany
    \\\{sandeep.singh, w.bziuk, a.jukan\}@tu-bs.de}
}

\maketitle

\begin{abstract}
Data randomization or scrambling has been effectively used  in various applications to improve the data security. In this paper, we use the idea of data randomization to proactively randomize the spectrum (re)allocation to improve connections' security. As it is well-known that random (re)allocation fragments the spectrum and thus increases blocking in elastic optical networks, we analyze the tradeoff between system performance and security. To this end, in addition to spectrum randomization, we utilize an on-demand defragmentation scheme every time a request is blocked due to the spectrum fragmentation. We model the occupancy pattern of an elastic optical link (EOL) using a multi-class continuous-time Markov chain (CTMC) under the random-fit spectrum allocation method. Numerical results show that although both the blocking and security can be improved for a particular so-called randomization process (RP) arrival rate, while with the increase in RP arrival rate the connections' security improves at the cost of the increase in overall blocking.

\end{abstract}

\section{Introduction}
\par Securing high data rate applications in optical networks against physical layer attacks or \emph{unauthorized observation} has long been subject of intense studies \cite{medard1998attack,fok2011optical}.  In elastic optical networks (EONs), the security challenge can be mitigated in various ways, among others by data scrambling along the code, time and frequency domains\cite{liu2014physical}.  The data scrambling along the multiple dimensions helps to resist the brute-force attacks, and makes it difficult for the attacker to decode the data. With the flexibility in assigning subcarriers, we believe that spectrum allocation in EONs also present a unique opportunity to provide optical layer security. If the spectrum randomization process is regularly performed, then only a portion of a particular user's data will be observed over a range of frequencies that is listened by an eavesdropper.

\par However, the random spectrum (re)reallocation fragments the spectrum, and as many previous studies have shown, increases the blocking in EONs \cite{shi2013effect,singh2016RaaS}. \cite{shi2013effect}  showed that the blocking probability due to bandwidth fragmentation in EONs depends on the size of the available spectrum blocks on a link, and  their alignment over different links on the routing paths. In \cite{singh2016RaaS}, we analytically showed that spectrum reallocation increases the overall blocking in an elastic optical link (EOL). To alleviate this problem, spectrum defragmentation scheme can be utilized to consolidate the free spectrum. During defragmentation, also some connections could be disrupted due to the retuning of transceivers and reconfiguration of optical switches. Although the disruption can be minimized using some techniques like ``make-before-break"  \cite{takagi2011disruption} or ``hitless" defragmentation methods  \cite{cugini2013push,proietti2012rapid}, spectrum reallocation is not desirable. Therefore, finding a tradeoff between random spectrum ``scrambling" and the need to defragment the spectrum to improve the performance is no easy task, which has not been addressed to date.  As we show, modeling a single EOL spectrum (re)allocation is analytically a challenge under the spectrum contiguity constraint, and it gets more complicated in networks due to an additional spectrum continuity constraint (assuming no spectrum converter in EONs)\cite{christodoulopoulos2013time,yu2013first,beyranvand2014analytical,rosa2015statistical,singh2016DaaS}.

\par In this paper, we scramble/randomize the spectrum usage pattern to secure elastic optical links, which we refer to as randomization-as-a-service (RaaS) scheme. The RaaS can be performed proactively at random or periodic intervals, with mean RP interarrival time ($\frac{1}{\lambda_S}$). However, we also show that blocking gets worse  due to this randomization process ($\propto$ $\lambda_S$). To this end, we utilize a defragmentation-as-a-service (DaaS) scheme, which is triggered every time a connection would be blocked because the spectrum is scrambled and fragmented. For the analytical modeling of the combined DaaS and RaaS, we model an EOL using the multi-class continuous-time Markov chain (CTMC)  and include RaaS and DaaS states in addition to regular data service states. Combining proactive RaaS and on-demand DaaS, we show that security is as good as in the RaaS system, while blocking is better than if only a RaaS system used, with much lower reconfiguration time (than call holding times) and for a range of moderate loads.


\par The rest of this paper is organized as follows. Section II presents the model description. In Section III, we present the overall blocking and the security analysis. Section IV evaluates the performance. Finally, we conclude the paper in Section V.

\section{Model Description}
\begin{figure}[t]
 \centering
\includegraphics[width= 0.45\textwidth, height=2.5 cm]{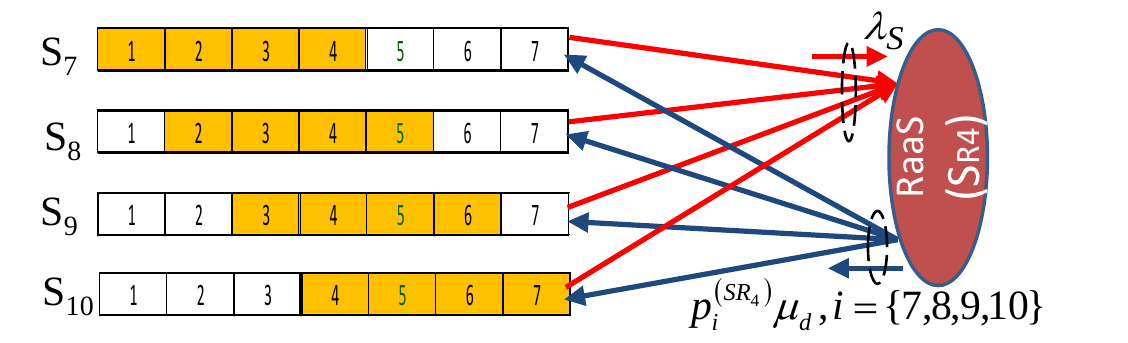}
\vspace{-0.2 cm}
  \caption{All possible state transitions from and into a RaaS state due to RP.}
 \vspace{-0.3cm}
\label{fig:spectrumResort}
\end{figure}

\begin{figure}[t]
 \centering
\includegraphics[width= 0.48\textwidth, height=2.6 cm]{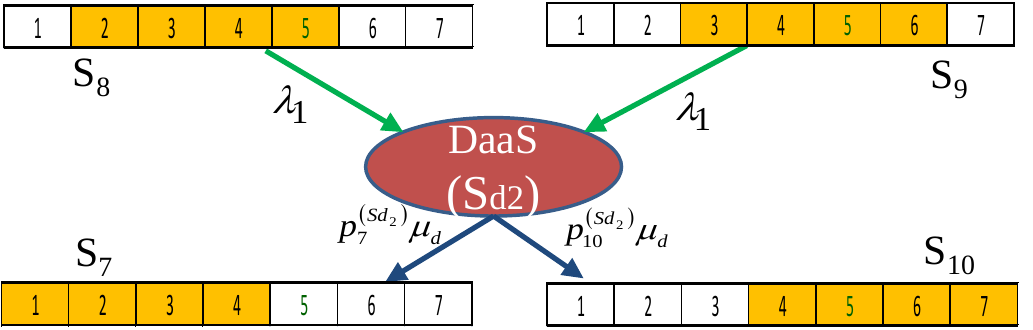}
\vspace{-0.2 cm}
  \caption{State transitions from and into a DaaS state for a demand of 3-slots.}
  \vspace{-0.3 cm}
\label{fig:defragTransitionExample}
\end{figure}
\par  The randomization of spectrum allocation makes it more difficult for an attacker to detect the spectrum assignment pattern and to successfully demodulate the spectrum, thus increasing the system security. At the same time, the randomization of spectrum allocation might result in fragmented spectrum. This is illustrated in Fig. \ref{fig:spectrumResort}, where the assignment of spectrum slots to existing connections is randomly reconfigured by entering the RaaS process (state $SR_4$), and as a result, one of the states $S_{7}, S_{8}, S_{9}$ and $S_{10}$ is chosen with equal probability.  Notably, the randomization feature can also be used to reallocate spectrum resources to consolidate free slots, which we call as a DaaS. This is illustrated in Fig. \ref{fig:defragTransitionExample}, where a request (say $R_i$) with demand $d = 3$ slots will be blocked in fragmented states $S_8$ and $S_9$ in a regular or RaaS system. In the combined RaaS-DaaS model, however, the arrival of the request $R_i$ into such fragmented states will trigger DaaS process (state $Sd_2$), which reconfigures the existing connection(s) and finally, the system will move to defragmented states $S_{7}$ or $S_{10}$ with equal probability.

\subsection{Underlining Assumptions}
\begin{figure*}[t]
 \centering
\includegraphics[width= 0.85\textwidth]{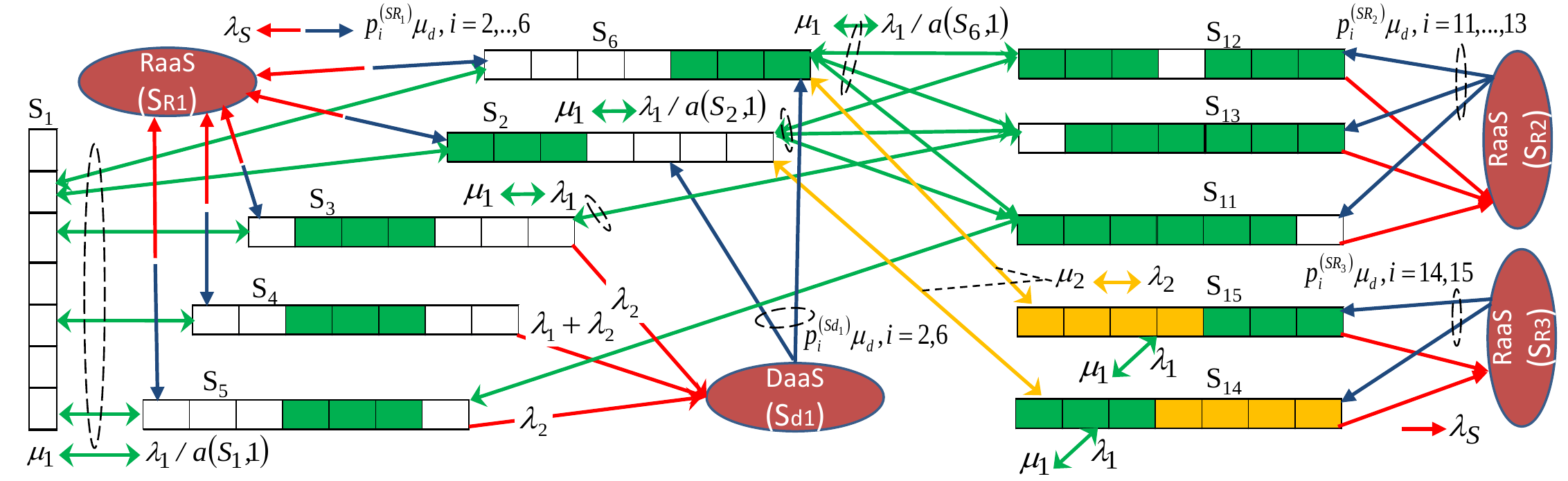}
\vspace{-0.3 cm}
  \caption{Partial state diagram with transitions leading to spectrum randomization and defragmentation in the RaaS-DaaS model under RF allocation policy.}
\vspace{-0.4 cm}
\label{fig:RFDefragExample}
\end{figure*}
\par We model the occupancy pattern of an EOL using a multi-class CTMC \cite{yu2013first,beyranvand2014analytical,rosa2015statistical,singh2016DaaS} with the following assumptions.
\newline i) The RaaS process is triggered for spectrum randomization at exponential RP inter-arrival times $T_{S}$ with average rate $\lambda_S$.
\newline ii) The DaaS process is triggered when a request is blocked due to the fragmentation of spectrum.
\newline iii)  Call inter-arrival times $T_{E_k}$  and call holding times $T_{H_k}$ of class-k requests are independent and  exponentially distributed with average rates $\lambda_k$ and  $\mu_k$, respectively.
\newline iv) In RaaS and DaaS states, reconfiguration times $T_{RT}$ are exponentially distributed with rate $\mu_d$.
\newline v) The services of all existing requests are interrupted during the reconfiguration times of the DaaS and RaaS processes.
\newline vi) Further incoming requests are blocked during the DaaS and RaaS periods, referred to as reconfiguration blocking.
\newline vii) We differentiate between resource blocking and blocking due to fragmentation; in the case of resource blocking, the total number of free slots does not satisfy the demand, while in the case of fragmentation blocking, there are enough free slots, but there is no sufficient amount of consecutive free blocks to satisfy the demand. 

\par Assumption (v) is disadvantageous, because traffic interruption of some or all existing connections does happen during reconfiguration in the real system, though there are efficient techniques to address this issue \cite{takagi2011disruption,cugini2013push,proietti2012rapid}.
Due to assumption (vi) we would experience an increase in overall call blocking probability. However, the results show that although reconfiguration blocking is added to the resource and fragmentation blocking (assumption vii), the overall blocking can be lower than a regular system without spectrum reallocation strategy, if the spectrum reallocation time is much lower than the normal service times of the connections.

\subsection{Mapping Transitions Between System States }\label{sec:state transition}
\begin{table}[t]
\caption{Notations and the parameters used in the model}
\vspace{-0.2cm}
 \centering
\scriptsize
  \begin{tabular}{@{}ll@{}}
\toprule
{\bf Notation} & {\bf Description} \\ \midrule
     $ C  $      &   Total number of spectrum slots (or capacity units)             \\ \midrule
     $ \lambda_k \,\, (\mu_k)  $      &    Arrival (service) rate of class $k$ calls, where  $k = 1, 2, \cdots, K$             \\ \midrule
     $ 1/\lambda_S $      &   Mean RP interarrival time ($E[T_S] = 1/\lambda_S$)          \\ \midrule
     $ 1/\mu_d$         &     Mean reconfiguration time  ($E[T_{RT}] = 1/\mu_d$)            \\ \midrule
     $\textbf{n}$      &     $ \equiv (n_1, n_2, \dots, n_K)$,  where $n_k$ is the number of class $k$ calls       \\ \midrule
     $S_i$          & Occupancy state for normal operation $i=1,...,N_{SA}$    \\ \midrule
     $Sd_{\nu}$     & Defragmentation state  $\nu=1,...,N_{D}$    \\ \midrule
     $SR_{\nu}$     & Randomization state  $\nu=1,...,N_{R}$    \\ \midrule
     $p_i^{Sd_\nu}$ ( $p_i^{SR_\nu}$) & transition probability from state $Sd_\nu$ ($SR_\nu$) to target state $S_i$ \\ \midrule
     $n_k(S_i)$     &     Number  of class-k calls in state $S_i$            \\ \midrule
     $\textbf{n}(S_i)$   & $ \equiv (n_1(S_i), \dots, n_K(S_i))$, realization of \textbf{n} in state $S_i$  \\ \midrule
     $a(S_i,k)$     &     Number  of different ways class-k call can be allocated in $S_i$            \\ \midrule
     $FS_l(S_i)$ & Size of $l^{th}$ fragment of free slots in state $S_i$ \\  \midrule
     $N_{SA}, N_{D}, N_{R} $ & Number of regular, DaaS, and RaaS states respectively \\
\bottomrule
\end{tabular}
\label{table:Notations}
\vspace{-0.5cm}
\end{table}

\par For the combined RaaS-DaaS model, some of the notations and parameters are listed and described in Table \ref{table:Notations}. Here, in addition to regular data service states ($S_i$), we have RaaS and DaaS states, which are used for spectrum randomization and defragmentation operations, respectively. Each RaaS (DaaS) state is associated with a non-overlapping set of states ($S_i$) with same connection pattern as $SR_\nu$ $(Sd_\nu)$, where a connection pattern is defined by the number of connections per class, i.e. $\textbf{n}(S_j)= (n_1(S_j),...,n_K(S_j))$. In general, we define a set of states with same connection pattern as $S_j$ as follows.
\begin{equation}
 \Gamma(S_j)=   \{ S_i |   \textbf{n}(S_i) = \textbf{n}(S_j), i = 1,...,N_{SA} \}
\label{eqn:GSj}
\end{equation}

\par It should be noted that a specific RaaS state $SR_\nu$ is triggered (i.e., $S_i \rightarrow SR_\nu$) by states, with same connection pattern but different spectrum occupancy pattern, belonging to the set $ \Gamma(SR_\nu)$. Furthermore, after randomization in the RaaS state $SR_\nu$, the transition $SR_\nu \rightarrow S_i$ can also be given by the set $\Gamma(SR_\nu)$. For example, in Fig. \ref{fig:spectrumResort}, all states out of set $ \Gamma(SR_4)=\{S_{7}, S_{8}, S_{9}, S_{10}\}$ transit to a RaaS state $SR_{4}$ at rate $\lambda_S$. And, after randomization process in $SR_\nu$, it transits back  to a randomly chosen system state $S_i \in  \Gamma(SR_4)$ with rate $p_i^{SR_4}\mu_d =\mu_d/4, i = \{7, 8, 9, 10\}$. Due to the randomization strategy, we assume equal transition probabilities $p_{i}^{SR_\nu}=\frac{1}{|\Gamma(SR_\nu)|}$, where $| \Gamma(SR_\nu|$ is the number of elements in $ \Gamma(SR_\nu)$. 

The mapping between a DaaS state to regular states, on the other hand, is not straightforward, since not all but only fragmented regular states $S_i$ with $\textbf{n}(S_i) = \textbf{n}(Sd_\nu)$ trigger transitions into the DaaS state $Sd_\nu$. Similarly, from $Sd_\nu$ transitions occur to only defragmented states $S_j$ satisfying $\textbf{n}(S_j) = \textbf{n}(Sd_\nu)$, where in a defragmented state $S_j$, all empty slots form a single block of free spectrum. Let us first define a fragmented  state, which can not allocate a large enough block $FS_l(S_i)$ of $l$ consecutive free slots to a class-k request of demand $d_k>l$, even though it contains equal or more than $d_k$ number of  free slots, as follows. 
\begin{equation}
fr(S_i,k) =
\begin{cases}
1 & \text{if} \, d_k > max_l FS_l(S_i) \,\text{and}  \\
   &  d_k \leq C- \sum_jd_jn_j(S_i) \\
0 & \text{otherwise}
\end{cases}
\end{equation}
Using the above function, the set of fragmented states for class-k requests is defined as
\begin{equation}
\mathbb{FB}(k) = \{ S_i | fr(S_i,k) =1, i=1, \cdots, N_{SA}\}, \forall k.
\end{equation}
Furthermore, the set of classes for which state $S_i$ is a fragmentation state is $\mathbb{FI}(S_i)=\{  k \mid S_i \in \mathbb{FB}(k) \},  i=1,\cdots,N_{SA}\}$.
Resource blocking states, on the other hand, do not have sufficient free slots to satisfy an incoming class-k request, i.e.,
\begin{equation}
\mathbb{RB}(k) = \{ S_i |  d_k > C- \sum\mathop{}_{\mkern-5mu j}d_jn_j(S_i), i=1,..., N_{SA}\}.
\end{equation}

Now using Eq. \eqref{eqn:GSj}, $ \Gamma(Sd_\nu)$ comprises all states with connection pattern defined by DaaS state $Sd_\nu$. The subset $\mathbb{FF}(Sd_\nu,k) =  \Gamma(Sd_\nu) \bigcap \mathbb{FB}(k)$ defines the set of states to be de-fragmented due to an arriving class-$k$ request, e.g. in Fig.\ref{fig:defragTransitionExample}, for a class-1 arrival the set $\mathbb{FF}(Sd_2,1)$ includes the fragmented states $S_8$ and $S_9$. Similarly, after reconfiguration, a DaaS state $Sd_\nu$ will transit to a new target state $S_i \in \mathbb{FT}(Sd_\nu)$ with probability $p_i^{Sd_\nu}$, e.g. in Fig.\ref{fig:defragTransitionExample} from state $Sd_2$ to state $S_{10}$ with rate $\mu_d p_{10}^{Sd_2}$. A target state is defined by the property, that all its free spectrum slots build one consecutive block $FS_1(S_i)$ irrespective of its spectral location. Thus, the set of target states is given as follows.
\begin{equation}
\mathbb{FT}(Sd_\nu) \!\! = \!\! \{ S_i | FS_1(S_i) \! = \! C \! -  \!\! \sum\mathop{}_{\mkern-5mu j} \!\!\!\! d_j n_j(S_i),\!\forall S_i \! \in \!  \Gamma(Sd_\nu)\}
\end{equation}
Hence, the  transition probability is given by $p_i^{Sd_\nu}=\frac{1}{|\mathbb{FT}(Sd_\nu)|}$. In Fig. \ref{fig:defragTransitionExample}, we have $\mathbb{FT}(Sd_2) = \{S_7, S_{10}\}$ with two possible target states, hence $|\mathbb{FT}(Sd_2)|=2$ and $p_{7}^{Sd_2}$ = $p_{10}^{Sd_2}$ = 1/2.


\par Now, we consider an example of an EOL with capacity $C=7$  slots under random-fit (RF) policy, see Fig. \ref{fig:RFDefragExample}. There are two different classes of requests with demands $3$ and $4$ consecutive slots. Here, we show only a part of all states and transitions, where a new call of class $k=1$ arrives in state $S_1$ with demand $d_1=3$. Under RF policy, it allocates one of the possible assignments shown by the transitions from $S_1$ to states $ \{S_2, S_3, S_4, S_5, S_{6}$\}. In the target states, we have $n_1(S_i)=1, i=2,...,6$ class-1 calls, which will leave $F_1=4$ free slots, thus there are $a(S_1,k=1)=\frac{(n_1+F_1)!}{n_1! F_1!}=5$ different ways to allocate the $7$ free slots, and transitions occur with rate $\lambda_1/5$. Similarly, a new call of class $k=2$ arrives in state $S_6$ with demand $d_2=4$ and can occupy only the possible state $S_{15}$. On the other hand, transition rates out of states $S_i, i=1,...,15$, will occur due to the departure of a class-1 and/or class-2 requests with rate $\mu_k, k=1,2$.
As we can see, when a class-2 request (4 slots) arrives, blocking of the request due to fragmentation should occur in the set of states $\mathbb{FB}(2)=\{S_3, S_4, S_5\}$. In this simple example, we have $\mathbb{FF}(Sd_1,2)=  \Gamma(Sd_1) \cap \mathbb{FB}(2) \equiv \mathbb{FB}(2) $, and the arrivals of such class-2 requests with rate $\lambda_2$ will trigger the transition from states $S_i \in \mathbb{FF}(Sd_1,2)$ to DaaS state $Sd_1$.
Similarly, the set of states $\mathbb{FB}(1)=\{S_4, S_8, S_9 \}$ are fragmented states for a class-1 call handled by DaaS states $Sd_1$ and $Sd_2$. Notice that we have $\mathbb{FF}(Sd_1,1)=  \Gamma(Sd_1) \cap \mathbb{FB}(1) = \{S_4\}$, and $S_4$ is also handled by $Sd_1$ for a class-1 arrival with rate $\lambda_1$. In total, this results into the transition from $S_4$ to $Sd_1$ with joint rate $\lambda_1+\lambda_2$. The remaining states out of $\mathbb{FB}(1)$  are handled by $Sd_2$ due to the set $\mathbb{FF}(Sd_2,1)= \Gamma(Sd_2) \cap \mathbb{FB}(1) = \{S_8,S_9\}$ as shown in Fig.\ref{fig:defragTransitionExample}.
In a similar form, all states out of the set $ \Gamma(SR_1)= \{S_2, S_3, S_4, S_5, S_{6}\}$ will also transit to a RaaS state $SR_1$, and because $| \Gamma(SR_1)|=5$ the process transits back to a randomly chosen state $S_i \in  \Gamma(SR_1)$ with rate $\mu_d/5$.

It should be noted that while the process is in DaaS or RaaS states, existing connections effected by spectrum reconfiguration are interrupted for a time duration equal to the exponentially distributed reconfiguration time $T_{RT}$ with mean $1/\mu_d$. Due to the memoryless property of the exponential distribution, the remaining call holding times of interrupted calls are again exponentially distributed with mean $1/\mu_k$. Thus, all transitions due to call terminations after interruption are also modeled with rates $\mu_k$.

\section{Blocking and Security Analysis} 

At first, we show how to derive the global balance equations (GBEs) of the Markov chain. Furthermore,  we define the following terms to calculate the transition rates of the system.
\begin{equation}
A(S_i,k) = \begin{cases}
1 & \text{if a class-k request is accepted, or} \\
   & \text{blocked due to fragmentation in}\, S_i \\
0 & \text{otherwise}
\end{cases}
\end{equation}
$A(S_i,k)$ determines if a transition from state $S_i$ is possible due to the arrival of a class-k call, and include the case that  $S_i$ is a fragmented state for this class, i.e $k \in  \mathbb{FI}(S_i)$. Additionally, $T^{\pm}(S_j, S_i, k)$ determines the cause of transition if it is due to allocation (+) or deallocation (-), given as
\begin{equation}
T^{\pm}(S_j, S_i, k) =
\begin{cases}
1 & \text{if } S_j \rightarrow  S_i  \text{ due to class-k} \\
  & \text{arrival (departure) in } S_j  \\
0 & \text{otherwise.}
\end{cases}
\end{equation}

To model the different many-to-one and one-to-many mapping between RaaS and regular states  $S_i$, we introduce an indicator function, which compares the connection pattern of states. 
\begin{equation}
\delta(S_i, SR_\nu) =
\begin{cases}
1 & \text{if } \textbf{n}(S_i)  = \textbf{n}(SR_\nu)  \\
0 & \text{otherwise}.
\end{cases}
\end{equation}
The above relation can also be used to define the many-to-one mapping from fragmented states $S_i$ to a DaaS state $Sd_\nu$ (i.e., $\delta(S_i, Sd_1)=1$ with the restriction $S_i \in \bigcup_k \mathbb{FF}(Sd_1,k)$. Similarly,  one-to-many mapping from a DaaS state $Sd_\nu$ to target defragmented states (in $\mathbb{FT}(Sd_\nu)$) is defined as follows.
\begin{equation}
\sigma(Sd_\nu, S_j) =
\begin{cases}
1 & \text{if }  S_j \in \mathbb{FT}(Sd_\nu)   \\
0 & \text{otherwise}
\end{cases}
\end{equation}


\par Finally, the GBE of each regular occupancy state $S_i$ under the RF spectrum allocation policy can be obtained by Eq. \eqref{eqn:GBEnormal}.
\begin{multline}
\left(\sum_{k=1}^{K}A(S_i,k)  \lambda_k  + n_k(S_i)\mu_k  + \lambda_S  \right ) \pi(S_i) = \\
\sum_{j=1, j\neq i}^{N_{SA}}  \sum_{k=1}^{K}   \left( \frac{\lambda_k T^+(S_j,S_i,k)}{a(S_j,k)} +  T^-(S_j,S_i,k) \mu_k\right) \pi(S_j) \\
+ \sum_{\nu=1}^{N_{D}} \sigma\left(Sd_\nu, S_i\right)\frac{\mu_d\pi(Sd_{\nu})}{|\mathbb{FT}(Sd_\nu)|}    \\
+  \sum_{\nu=1}^{N_{R}}   \delta\left(SR_{\nu}, S_i\right)   \frac{ \mu_d  \pi(SR_{\nu}) }{| \Gamma(SR_\nu)|}, i =1, \cdots, N_{SA}
\label{eqn:GBEnormal}
\end{multline}

In Eq. \eqref{eqn:GBEnormal}, left hand side represents the output flow rate from the state $S_i$ including transitions to DaaS (taken into account by $A(S_i,k)$) and RaaS states (rate $\lambda_S$), while the right hand side represents input flow rate into the sate $S_i$. More precisely, the second line of Eq. \eqref{eqn:GBEnormal} represents the input flows from other regular states $S_j$, while the third line of Eq.\eqref{eqn:GBEnormal} defines the rate from exactly one DaaS state to state $S_i$. 
The indicator function $\sigma(Sd_\nu, S_i) =1$ selects a DaaS state only if $S_i \in \mathbb{FT}(Sd_\nu)$ i.e., if $S_i$ is a target defragmented state, and it can be reached after reconfiguration in $Sd_\nu$ with rate $\frac{\mu_d}{|\mathbb{FT}(Sd_\nu)|}$.
In the last line, the indicator $\delta\left(SR_\nu), S_i\right)$ selects a correct RaaS state $SR\nu$ with connection pattern equivalent to the state $S_i$, where the factor $| \Gamma(SR_\nu)|$ takes into account, that the RaaS state only randomly selects a target state $S_i$ with the probability $1/| \Gamma(SR_\nu)|$.

Similarly, the GBEs of a RaaS state $SR_\nu$ and a DaaS state $Sd_\nu$ can be given by Eq. \eqref{eqn:GBEreconfig} and Eq. \eqref{eqn:GBEdefrag}, respectively.
\begin{equation}
\mu_d \pi(SR_{\nu})=   \lambda_S    \sum_{S_j \in  \Gamma(SR_\nu)}\!\!\!\!\!\pi(S_j),  \nu = 1, 2, \cdots, N_R
 \label{eqn:GBEreconfig}
\end{equation}
\begin{equation}
\mu_d\pi(Sd_{\nu})\!\!=\!\!\sum_{k=1}^{K}\!\lambda_k \!\!\!\!\sum_{S_j \in \mathbb{FB}(k)}\!\!\!\!\!\!\!\delta\left(S_j,Sd_{\nu}\right)\pi(S_j),\nu = 1,..., N_D
 \label{eqn:GBEdefrag}
\end{equation}
In Eq. \eqref{eqn:GBEdefrag}, the function $\delta\left( S_j, Sd_{\nu} \right)$ selects the corresponding fragmented states $S_j \in \mathbb{FB}(k)$ with related rate $\lambda_k$.

As an example, the GBE for state $S_4$  in Fig.\ref{fig:RFDefragExample} (which is the only one of the fragmentation states for both classes) is given by $(\lambda_1 + \lambda_2 +  \mu_1 + \lambda_S)\pi(S_4) = \lambda_1/5 \pi(S_1) + \mu_d/5 \pi(SR_1)$. 
The input rate $\lambda_1/5$ ($\mu_d/5$) is due to transition $S_1 \rightarrow S_4$ ($SR_1 \rightarrow S_4$), as explained in Sec. \ref{sec:state transition}. 
For the output rate, a departure of an existing class-1 connection takes place with rate $\mu_1$, and RP requests arrive with rate $\lambda_S$. Furthermore, we have two different classes of requests that will be blocked due to fragmentation, i.e. $S_4 \in \mathbb{FF}(Sd_1,1) $ and $S_4 \in \mathbb{FF}(Sd_1,2) $, and thus will cause transition to a DaaS state $Sd_1$ with total rate $\lambda_1+\lambda_2$. Similarly, we can write the GBE for a RaaS state $SR_3$, which is reached from a set of states $ \Gamma(SR_3) = \{S_{14}, S_{15}\}$ when a RP request arrives with rate $\lambda_S$ in one of these states. After randomization in $SR_3$, system will return back to one of the possible states in $ \Gamma(SR_3)$. However, the total output flow rate from $SR_3$  is still $\mu_d$. Hence, the GBE of state $SR_3$ is given by $\mu_d \pi(SR_3)= \lambda_S\left(\pi(S_{14}) + \pi(S_{15})\right)$. The GBE of a DaaS state $Sd_1$ can also be written as $\mu_d \pi(Sd_1)= \lambda_2\pi(S_3) + (\lambda_1+\lambda_2) \pi(S_4) + \lambda_2\pi(S_5)$. In this example, we have $\delta(S_j, Sd_1)=1 $ for regular state $S_j \in \mathbb{FB}(1)$, which selects a state $S_4$ for class-1, as well as $S_j \in \mathbb{FB}(2)$ which selects states ${S_3,S_4,S_5}$ for class 2 arrivals.

Under the stationary condition, the state probabilities  $\pi = [\pi(S_1), \pi(S_2), \cdots, \pi(S_{N_{SA}}), \pi(SR_1), \cdots, \pi(SR_{N_R} ),$ $\pi(Sd_1), \cdots, \pi(Sd_{N_D})  ]$ can be calculated by solving $\pi$ $Q=0$ subject to $\sum_i \pi(S_i) + \sum_\nu \pi(SR_\nu) + \sum_\nu \pi(Sd_\nu)= 1 $, where $N=N_{SA}+N_R +N_D$ is the total number of states, and $Q$ is the transition rate matrix.


The blocking per class-k due to the unavailability of resources, i.e., resource blocking (RB),  and due to fragmentation, i.e. fragmentation blocking (FB) can be calculated using Eq. \eqref{eqn:resourceBlocking} and Eq. \eqref{eqn:fragmentBlocking}, respectively.
\noindent\begin{tabularx}{0.49\textwidth}{@{}XX@{}}
\vspace{-0.5cm}
\begin{equation}
  RB(k)\!\!=\!\!\!\!\!\sum_{S_i\in \mathbb{RB}(k)}\!\!\!\!\!\pi(S_i)
\label{eqn:resourceBlocking}
  \end{equation} &
  \vspace{-0.5cm}
  \begin{equation}
FB(k)\!\!=\!\!\!\!\!\sum_{S_i\in \mathbb{FB}(k)}\!\!\!\!\!\pi(S_i)
\label{eqn:fragmentBlocking}
  \end{equation} 
\end{tabularx}
In a RaaS system (I), i.e., without DaaS states and associated transitions, reconfiguration blocking can be given as follows.
\begin{equation}
RCB^I =\sum_{\nu=1}^{N_R}\pi(SR_{\nu})
\label{eqn:reconfigBlocking}
\end{equation}
On the other hand, the reconfiguration blocking in the combined RaaS-DaaS system (II) is given as follows.
\begin{equation}
RCB^{II} =\sum_{\nu=1}^{N_R}\pi(SR_{\nu})+\sum_{\nu=1}^{N_D}\pi(Sd_{\nu})
\label{eqn:defragBlocking}
\end{equation}
It should noted that RaaS and DaaS states are blocking states for all classes of requests.
Finally, the overall blocking in the RaaS system (I) and the combined RaaS-DaaS system (II) can be given by the Eq. \eqref{eqn:RaaSBlocking} and Eq. \eqref{eqn:DaaSBlocking}, respectively.
\begin{equation}
BP^{I} = RCB^{I}  + \frac{\sum_k \lambda_k ( RB(k) + FB(k) )}{\sum_k\lambda_k}
\label{eqn:RaaSBlocking}
\end{equation}
\begin{equation}
BP^{II}  = RCB^{II}  + \frac{\sum_k \lambda_k ( RB(k) + FB(k) )}{\sum_k\lambda_k}
\label{eqn:DaaSBlocking}
\end{equation}

\par In a regular system i.e., without DaaS and RaaS states, the GBEs can be simplified. In Eq. \eqref{eqn:GBEnormal}, all transitions from and into DaaS ($Sd_\nu$) and RaaS ($SR_\nu$) states are not present, which can simply be modeled by setting $\mu_d = 0$ and $\lambda_S = 0$.  Furthermore, Eq. \eqref{eqn:GBEdefrag} and \eqref{eqn:GBEreconfig} have to be omitted. In the regular system, in additional to the blocking due non-availability of resources, blocking also occurs due to the fragmentation. Hence, the overall blocking for regular system can be calculated as, i.e.,
\begin{equation}
BP^{regular} = \frac{\sum_k \lambda_k ( RB(k) + FB(k) )}{\sum_k\lambda_k}
\label{eqn:regularBlocking}
\end{equation}

\subsection{Security Analysis}
\begin{figure}[t]
 \centering
\includegraphics[width= 0.45\textwidth, height=1 cm]{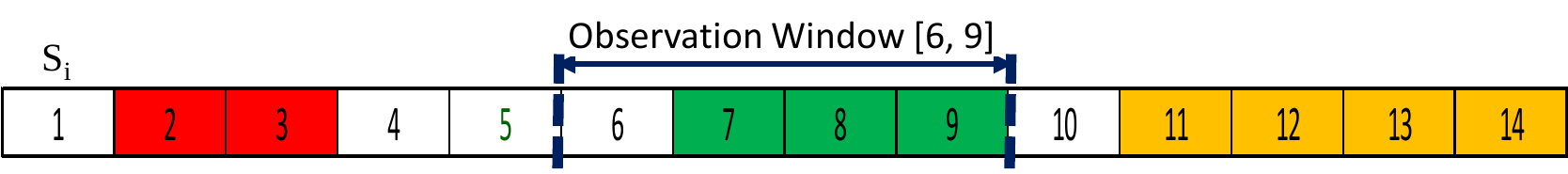}
\vspace{-0.2 cm}
  \caption{An example of eavesdropping in spectrum window [6, 9].}
 \vspace{-0.4 cm}
\label{fig:observationWindow}
\end{figure}
\par In this subsection, we show how spectrum randomization can improve the connections' security. Consider a scenario, where an eavesdropper gets access to a transponder with fixed capacity $f_0 - f_n$ to spoof the data of a particular user. It should be noted that if RaaS is not performed within the lifetime of the user's connection allocated over $f_0 - f_n$, then all data volume of the connection is observed. Therefore, in this case, security of those data volumes depends if the eavesdropper can decode the  data or not. However, in our model, only a fraction of the user's data will be observed by the attacker if RaaS is performed frequently, which enhances the data security on top of the encryption. RaaS states are included for this purpose, which randomly reassign subcarrier slots to connections. We assume that RaaS is triggered at exponential RP inter-arrival times $T_{S}$ with average rate $\lambda_S$. And, after randomization, it returns to one of all possible states out of set $ \Gamma(SR_\nu)$ having same connection pattern (but different occupancy patterns) as shown in Fig. \ref{fig:spectrumResort} and Fig. \ref{fig:RFDefragExample}. Let us calculate the total number of possible rearrangements of connections in a state $S_i$. If the state $S_i$ contains $n_k(S_i)$ connections of $class-k$ with demand $d_k$, and there are $E(S_i) = C - \sum_{k=1}^K n_k(S_i) d_k$ empty slots. Then, for the RF policy and without distinguishing the same class of connections, the number of all possible rearrangements of connections of $S_i$ with spectrum contiguity constraint can be given as follows \cite{beyranvand2014analytical}.
\begin{equation}
 R_{\textbf{n}}(S_i) = \frac{\left(E(S_i) + \sum_{k=1}^K n_k(S_i)\right)!}{E(S_i)! \prod_{k=1}^K n_k(S_i)!}
\label{eqn:TotalRearrangement}
\end{equation}

\par Now, let us assume that an eavesdropper can observe any part of spectrum with slot-width $W$, where $W \leq C$, and the observed window is uniformly distributed over the entire spectrum. For example in Fig. \ref{fig:observationWindow}, the probability ($p^W$) that at any point in time the observed window lies in the range [6, 9] is 1/11, since there are 11 (=14 - 4 +1) different ways of observation across spectrum with $C = 14$ and $W = 4$. In general, for a uniform distribution, $p^W =\frac{1}{C-W+1}$. Furthermore, an attack (eavesdrop) is defined as successful if the observed connections within the spectrum window ($\mathbf{n_{in}}$) before and after a randomization process remains same, and the probability of a successful attack for an observation window of size $W$ (i.e., $P^{W}_{SA}$) is given as follows.
\begin{multline}
P^{W}_{SA} = \sum_{i =2}^{N_{SA}}\left[P^W_{SA}(S_i) \times  \frac{\pi(S_i)}{\sum_{j=2}^{N_{SA}}\pi(S_j)} \right], \\
\text{where} \,\, P^W_{SA}(S_i) = \sum_{j=1}^{C-W+1}p^W\times\frac{R^j_{\mathbf{n_W}}(S_i)}{R_{\mathbf{n}}(S_i)}
\label{eqn:SAP}
\end{multline}

\par In Eq. \eqref{eqn:SAP}, $P^W_{SA}(S_i)$ is the successful attack probability given that the system is in state $S_i$, and it is given by the expected value of the ratio of $R^j_{\mathbf{n_W}}(S_i)$ and $R_{\mathbf{n}}(S_i)$. This ratio gives the successful attack probability for a given window $[j, j+W-1]$ and state $S_i$. Here,  $R^j_{\mathbf{n_W}}(S_i)$ defines the possible number of rearrangements of connections of $\textbf{n}(S_i)$ which results in same $\mathbf{n_{in}}$ in the window $[j, j+W-1]$ before and after the randomization. Consider  example in Fig. \ref{fig:observationWindow}, where an attacker listens to a connection (with demand 3-slots) within its observation window [6, 9] when the system is in state $S_i$. Here, there is a single connection for every class with demand $d_k =\{2, 3, 4\}, k=1,2$ and 3. Therefore,  for a given connection pattern $\mathbf{n}(S_i) = (1, 1, 1)$ and 5 empty slots, the randomization process will result in one of all possible rearrangements i.e., 280 $(= \frac{(5 + 3)!}{5!})$ . It should be noted that there is only one connection with $d_k = 3$-slots inside the observation window. Therefore, for an attack to be successful, the inside connection pattern $\mathbf{n_{in}}(S_i) = (0, 1, 0)$ should be same as before and after the randomization. In general, the number of such rearrangements ($R^j_{\mathbf{n_W}}(S_i)$) is not straightforward. However, for a small-scale EOL, we can calculate the number of such rearrangements by counting all states that comprise same connection pattern $\mathbf{n}$, and same $\mathbf{n_{in}}$. 

\begin{figure*}[t]
 \centering
\includegraphics[width=0.95\textwidth, height=2.2in]{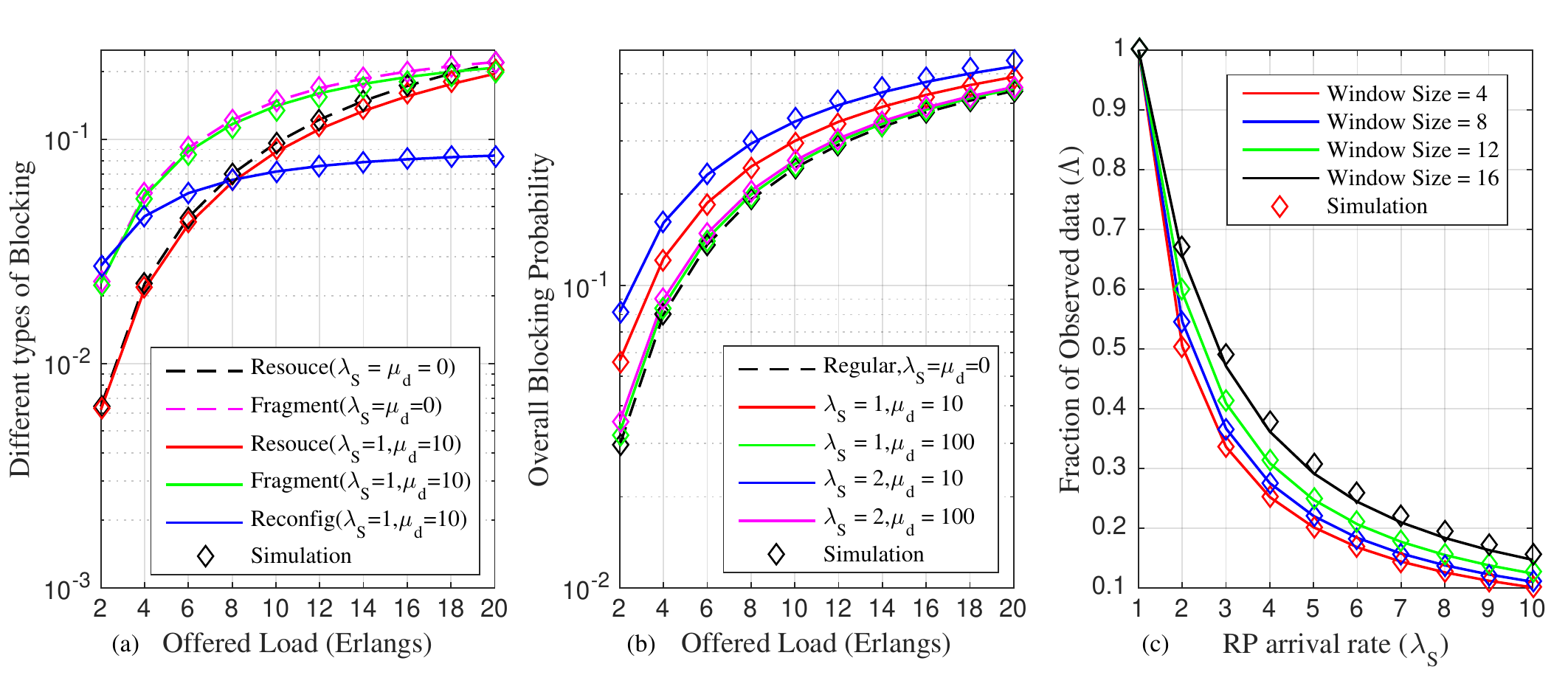}
  \vspace{-0.3 cm}
  \caption{Evaluation of RaaS model with C = 20: (a) blocking parts in a Regular and RaaS system; (b) overall blocking in a Regular and RaaS models for various reconfiguration rate ($\mu_d$) and RP arrival rates $\lambda_S$; and (c) fraction of observed data in RaaS model for various $\lambda_S$ at a fixed $\mu_d = 100$.}
\label{fig:BlockingAll}
\vspace{-0.3 cm}
\end{figure*}

In Fig. \ref{fig:observationWindow}, it is easy to see that the number of rearrangements inside the window [6, 9] i.e.,  $R^6_{\mathbf{n_{in}}}(S_i) =  2$ (Eq. \eqref{eqn:TotalRearrangement}); and there are 16 possible rearrangements of connections and free slots outside the window (i.e., $R^6_{\mathbf{n_{out}}}(S_i)=16$, explained latter). Therefore, there are $R^6_{\mathbf{n_W}}(S_i)=32$  possible spectrum occupancy patterns out of 280, where a 3-slots connection within the window [6, 9] can be successfully observed even after the randomization. Now, we show in three steps how to find the number of rearrangements outside the window, which is similar to a ``partition" problem -- a NP hard problem\cite{hayes2002computing}.
Firstly, form a multiset $\mathbb{E}$ of elements as 1 (for each free slot) and demand $d_k$ (for connections in $\mathbf{n_{out}}(S_i)$). Secondly, find the number of possible ways of partitioning of $\mathbb{E}$ into two subsets $\mathbb{E}_1$ and $\mathbb{E}_2$ such that $\sum_{e_i\in\mathbb{E}_1}e_i = C_L$ and $\sum_{e_i\in\mathbb{E}_2}e_i = C_R$, where $C_L$ and $C_R$ are the number of slots to the left and to the right of the window, respectively. Finally, for each partition set find the possible permutations of the elements. In Fig. \ref{fig:observationWindow}, we have four free slots and two connections with demands $d_1=2$ and $d_3=4$ (i.e., $\mathbf{n_{out}}(S_i) = (1, 0, 1)$ for $K =3$) outside the window. Here,  $\mathbb{E}=\{1,1,1,1,2,4\}$, and $C_L=C_R=5$. There are two partitions \{$e_1 =(1,1,1,2), e_2=(1,4)$\} and \{$e_1 =(1,4), e_2=(1,1,1,2)$\} which sum to 5 each. Therefore, the total number of permutations of partition set elements is $2\times\frac{4}{3!}\times 2! = 16$,  thus $R^6_{\mathbf{n_{out}}}(S_i) = 16$. 

The average number of RP performed during a lifetime of a connection ($N_{r}$) can be calculated  as the ratio of the average holding time of a connection and the average RP interarrival time i.e.,   $N_{r} =\frac{1/\mu_k}{1/\lambda_S} = \frac{\lambda_S}{\mu_k}$, where we restrict $N_r \in \mathbb{N}$ for $\lambda_S = n\mu_k, n = \{1, 2, \cdots\}$.  Until RP is triggered for the first time, an attacker observes data successfully with probability 1. Therefore, the amount of data successfully observable before RP is $b \times \frac{1}{\lambda_S}\times 1$, where $b$ is the average data rate observed in a window of size $W$.
Similarly, until the $2^{nd}$ RP, successfully observable data is $b \times \frac{1}{\lambda_S}\times [1 + P^W_{SA}]$.
In general, the amount of successfully observable data until $N_r^{th}$ RP is given as follows.
\begin{multline}
D^W_{SA} =  b \times \frac{1}{\lambda_S} \left[ 1 +  {P^W_{SA}} + \cdots +  [{P^W_{SA}}]^{N_r - 1}  \right]  \\
             =  b \times \frac{1}{\lambda_S} \times \frac{1 - [{P^W_{SA}}]^{N_r}}{ 1 - P^W_{SA}}
\label{eqn:SAD}
\end{multline}
Now, the total amount of data transmitted during the mean holding time $1/\mu_k$  is $ \frac{b}{\mu_k}$. Therefore, the fraction of successfully observed data is given as the ratio of successful observed data and total transmitted data, and given as follows.
\begin{equation}
\Lambda^{[1,W]} =  \frac{\mu_k}{\lambda_S} \times \frac{1 - [{P^W_{SA}}]^{\frac{\lambda_S}{\mu_k}}}{ 1 - P^W_{SA}}
\label{eqn:FractionSAD}
\end{equation}

\section{Numerical Results}
\begin{table}[ht!]
\centering
\caption{Parameters used for the numerical results}
\vspace{-0.2 cm}
\label{table:parameters}
\begin{tabular}{@{}ccc@{}}
\toprule
EOL  capacity                                                             & C = 20      & C = 100       \\ \midrule
\begin{tabular}[c]{@{}c@{}}Demands ($d_k$)\end{tabular}          & \{4, 6, 8\} & \{5, 10, 15\} \\ 
\begin{tabular}[c]{@{}c@{}}Reconfiguration rate ($\mu_d$)\end{tabular} & \{10, 100\} & \{100, 1000\} \\ 
\begin{tabular}[c]{@{}c@{}}RP arrival rate ($\lambda_S$)\end{tabular} & \{1, 2,$\cdots$, 10\} & \{1, 2,$\cdots$, 10\} \\\bottomrule
\end{tabular}
\vspace{-0.2 cm}
\end{table}

\begin{figure*}[ht!]
 \centering
\includegraphics[width=0.95\textwidth, height=2.2in]{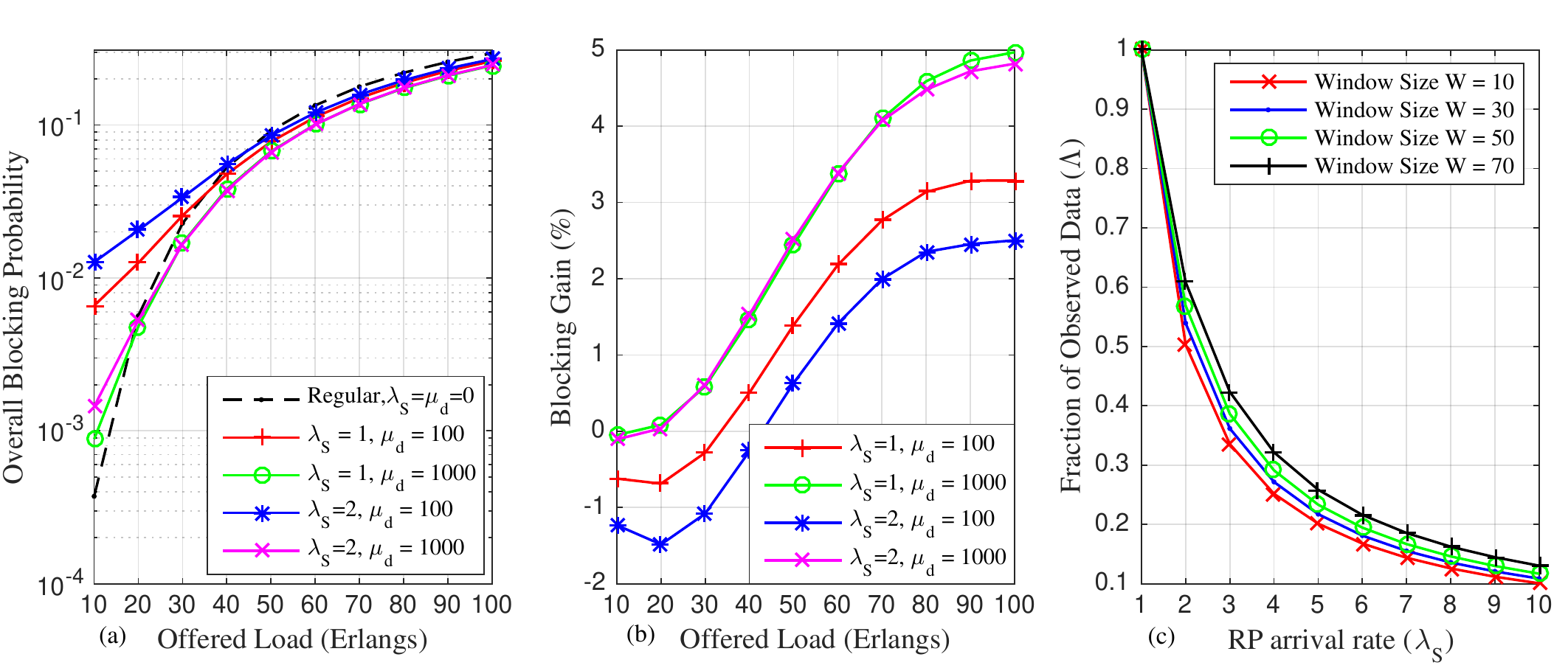}
  \vspace{-0.3 cm}
  \caption{Evaluation of the combined RaaS-DaaS model with C = 100: (a) overall blocking in a Regular and RaaS-DaaS models; (b) Blocking gain (in \%) in RaaS-DaaS model as compare to a Regular system; and (c) fraction of observed data in RaaS-DaaS model for various $\lambda_S$ at a fixed $\mu_d = 100$.}
\label{fig:BlockingAll100}
\vspace{-0.3 cm}
\end{figure*}

\par In this section, we present the analytical and verifying Monte Carlo (MC) simulation results for a small-scale EOL with capacity C = 20 spectrum slots. Since the number of states increases exponentially with the number of slots, we present MC simulation results for a large-scale EOL with capacity C = 100 slots. The parameters used for the two different scenarios are listed in Table \ref{table:parameters}. Arrival rates are uniformly distributed i.e., $\lambda_k = \frac{\lambda}{K}$,  where $\lambda$ is the total arrival rate. We assume mean holding time as one unit for all classes of requests \cite{yu2013first,beyranvand2014analytical,rosa2015statistical}. Load on the link is calculated as $\sum_{k =1 }^K\frac{\lambda_k d_k}{\mu_k}$.  We consider random-fit spectrum allocation for the performance evaluation due to the nature of our model which randomizes the spectrum irrespective of the spectrum allocation policy.

In Fig. \ref{fig:BlockingAll}, we evaluate our RaaS model (i.e., with RaaS states, but without DaaS states) against a regular system without spectrum reallocation.  Fig. \ref{fig:BlockingAll}a shows the blocking parts  due to the lack of resources (RB, Eq. \ref{eqn:resourceBlocking}), fragmentation (FB, Eq. \ref{eqn:fragmentBlocking}) and reconfiguration (RCB, Eq. \ref{eqn:reconfigBlocking}). As expected, in a regular and RaaS systems and for traffic load range shown here, the blocking is dominated by fragmentation and not by resource unavailability.  In Fig. \ref{fig:BlockingAll}b, blocking in our RaaS model is always higher than the regular system irrespective of the $\lambda_S$ and $\mu_d$. However, the overall blocking in RaaS system comes closer to the blocking in regular system when $\mu_d = 100$. The reason is that when mean reconfiguration time ($E[T_{RT}]=1/\mu_d$) reduces (i.e., rate is increased from 10 to 100), then the number of requests arriving during reconfiguration also reduces, and hence the RCB reduces. It should be noted that when the average RP interarrival time decreases (i.e., $\lambda_S$ increases), then blocking also increases significantly.
The reason is that RaaS states are called frequently and randomization does result in fragmented spectrum. However, with the increase in the average RP arrival rate ($\lambda_S$), the fraction of observed data decreases exponentially (in Fig. \ref{fig:BlockingAll}c), which means security also increases exponentially. Here, the fraction of observed data is plotted (Eq. \eqref{eqn:FractionSAD}) at fixed link load of 20 Erlangs. Additionally, we see that when the size of the observation window $W$ increases, the fraction of observed data also increases due to the increase in the probability of success (Eq. \eqref{eqn:SAP}). Note that when window size ($W$) equals to capacity ($C$), full spectrum can  be observed with probability 1 (Eq. \eqref{eqn:SAP}), and the attacker will observe 100\% of data.

It should be noted that RaaS model provides security at the cost of a slight increase in blocking. Therefore, now we present in Fig. \ref{fig:BlockingAll100} performance of the combined RaaS-DaaS model to show how the DaaS process impacts blocking and security for a large-scale EOL with capacity 100 slots.
As compare to regular system and  RaaS system, the overall blocking in our combined RaaS-DaaS model highly depends on the load (in Fig. \ref{fig:BlockingAll100}a). To understand this, we plot blocking gain against the load for various $\lambda_S$ and $\mu_d$ in Fig. \ref{fig:BlockingAll100}b. Here, blocking gain is defined as the change in blocking (in \%) in our combined model as compare to the regular system. First, we see that at lower loads blocking is higher irrespective of the $\lambda_S$ and $\mu_d$. This is due to the fact that $\lambda_S \gg \lambda_k$, and therefore the system spends more time in RaaS states which blocks all incoming requests. At moderate loads, blocking gain is positive and increasing with load. However, note that with the increase in $\lambda_S$, the load (connection arrival rate) should also increase to get the positive gain. Finally, at higher loads gain again decreases due to increase in fragmentation. At last, the security in the combined model (Fig. \ref{fig:BlockingAll100}c) is as good as in the RaaS model (Fig. \ref{fig:BlockingAll}c) due to the fact that defragmentation is not just shifting the spectrum usage towards one end, but also randomizes the connections while keeping free slots intact.

The most important findings of this paper are: i) in RaaS model, increase in blocking can be minimized as compared to the regular system without reconfiguration depending on the mean reconfiguration time ($E[T_{RT}]=1 / \mu_d$);  ii) however, if the mean reconfiguration time is much lower  (e.g. $\mu_d = 1000$) than the  mean holding time $E[T_{H_k}]$, the overall blocking in our combined RaaS-DaaS model is lower than the regular system without reconfiguration, for a range of moderate loads; and iii) the security can be improved by many folds with only a smaller RP arrival rate. Based on our findings, network operator can tune the reconfiguration times and RP arrival rate for a particular load value of the network in order to experience the higher gain from the our model.

\section{Conclusion}
In this paper, we examined the effect of spectrum reallocation on the connections blocking and the associated security. Based on our findings, we conclude that elastic optical network operator can tune the mean RP interarrival time (1/$\lambda_S$) and the mean reconfiguration time (1/$\mu_d$) for a particular load value of the network to experience the higher gain from our model. In our combined model, we showed the effect on the overall blocking probability of the system as function of the load, reconfiguration rate ($\mu_d$), the RP arrival rate ($\lambda_S$). Additionally, the fraction of observed data decreases with the RP arrival rate. For future work, we plan to extend our model to a multi-hop EONs. This is not an easy task, since again, both criteria of spectrum continuity as well as contiguity must be considered in the analysis which can make it rather complex.

\bibliographystyle{IEEEtran}
\bibliography{Defrag}

\end{document}